\documentclass[12pt]{article}
\usepackage{amsmath,amsfonts,amsthm,amssymb,amscd,latexsym, bm,bbm}
\usepackage{hyperref}
\usepackage[usenames,dvipsnames]{color}
\usepackage{authblk}
\hypersetup{colorlinks,
citecolor=Blue,
linkcolor=Blue,
urlcolor=Blue}

\usepackage[T2A]{fontenc}
\usepackage{graphicx}
\usepackage{mathrsfs}

{\theoremstyle {definition} \newtheorem {defi} {Definition} [section] }
{\theoremstyle {plain}  \newtheorem {theo} [defi] {Theorem}}
{\theoremstyle {plain}  \newtheorem {cor} [defi]{Corollary}}
{\theoremstyle {plain}  \newtheorem {lem} [defi]{Lemma}}
{\theoremstyle {plain}  \newtheorem {prop} [defi]{Proposition}}
{\theoremstyle {plain} \newtheorem {rem}[defi] {Remark}}
{\theoremstyle {definition} }

\numberwithin{equation}{section}
\newtheorem*{defi*}{Definition}
\newtheorem*{problem*}{Problem}

\newtheorem*{rem*}{Remark}
\newtheorem*{note*}{Note}

\makeatletter \@addtoreset{equation}{section} \makeatother

\newcommand{\mC}{\mathbb{C}}
\newcommand{\mR}{\mathbb{R}}

\newcommand{\mI}{\mathbb{I}}

\newcommand{\AAA}{{\cal A}}

\newcommand{\CC}{{\cal C}}
\newcommand{\DD}{{\cal D}}
\newcommand{\EE}{{\cal E}}
\newcommand{\FF}{{\cal F}}
\newcommand{\GG}{{\cal G}}

\newcommand{\LL}{{\cal L}}
\newcommand{\MM}{{\cal M}}
\newcommand{\NN}{{\cal N}}

\newcommand{\TT}{{\cal T}}

\newcommand{\eps}{\varepsilon}
\newcommand{\ph}{\varphi}
\newcommand{\thet}{\vartheta}

\newcommand{\ka}{\kappa}

\newcommand{\al}{\alpha}

\newcommand{\la}{\lambda}
\newcommand{\La}{\Lambda}
\newcommand{\de}{\delta}

\newcommand{\Ga}{\Gamma}

\newcommand{\dist}{\operatorname{dist}}

\newcommand{\ssupp}{\operatorname{sing\,supp}}
\newcommand{\supp}{\operatorname{supp}}

\newcommand{\intt}{\operatorname{int}}
\newcommand{\eff}{\operatorname{eff}}
\newcommand{\ov}{\overline}
\newcommand{\wt}{\widetilde}

\newcommand{\volna}{\thicksim}

\newcommand{\ssk}{\smallskip}

\newcommand{\p}{\partial}
\newcommand{\fr}{\frac}

\newcommand{\sm}{\setminus}

\newcommand{\ran}{\rangle}
\newcommand{\lan}{\langle}
\newcommand{\ds}{\displaystyle{}}

\newcommand{\ra}{\rightarrow}

\newcommand{\qu}{\quad}
\newcommand{\qmb}{\quad\mbox}
\newcommand{\qnd}{\quad\mbox{and}\quad}

\newcommand{\ass}{\quad\mbox{as}\quad}

\newcommand{\sli}{\sum\limits}
\newcommand{\ili}{\int\limits}
\newcommand{\ilif}{\ili_{-\infty}^\infty}
\newcommand{\intf}{\int_{-\infty}^\infty}

\newcommand{\non}{\nonumber}
\newcommand{\lbl}{\label}
\newcommand{\rprop}{Proposition \nolinebreak}
\newcommand{\rtheo}{Theorem \nolinebreak}
\newcommand{\rlem}{Lemma \nolinebreak}
\newcommand{\rcor}{Corollary \nolinebreak}

\newcommand{\rsec}{Section \nolinebreak}

\newcommand{\bee}{\begin{equation}}
\newcommand{\eee}{\end{equation}}
\newcommand{\bt}{\begin{theo}}
\newcommand{\et}{\end{theo}}
\newcommand{\bl}{\begin{lem}}
\newcommand{\el}{\end{lem}}
\newcommand{\bp}{\begin{prop}}
\newcommand{\ep}{\end{prop}}
\newcommand{\bc}{\begin{cor}}
\newcommand{\ec}{\end{cor}}
\newcommand{\bd}{\begin{def}}
\newcommand{\ed}{\end{def}}
\newcommand{\br}{\begin{rem}}
\newcommand{\er}{\end{rem}}

\title{Asymptotic behaviour of a network of oscillators coupled to thermostats of finite energy}
\author{Andrey V. Dymov\thanks{dymov@mi.ras.ru}}
\affil{Steklov Mathematical Institute of RAS, Moscow, Russia}
\date{}

\begin{document}
\maketitle

\begin{abstract}
We study the asymptotic behaviour of a finite network of oscillators (harmonic or anharmonic) coupled to a number of deterministic Lagrangian
thermostats of {\it finite} energy. In particular, we consider a chain of oscillators interacting with two thermostats situated at the boundary of the chain. 
Under appropriate assumptions we prove that the vector $(p,q)$ of moments and coordinates of the oscillators in the network satisfies $(p,q)(t)\to (0,q_c)$ when $t\to\infty$,
where $q_c$ is a critical point of some effective potential, so that the oscillators just stop. 
Moreover, we argue that the energy transport in the system stops as well without reaching the thermal equilibrium.
This result is in contrast to the situation when the energies of the thermostats are {\it infinite}, studied for a similar system in \cite{EPRB} and subsequent works, where the convergence to a non-trivial limiting regime was established.

The proof is based on a method developed in \cite{Tr},
where it was observed that
the thermostats produce some effective dissipation
despite the Lagrangian nature of the system.
\end{abstract}

\tableofcontents

\section{Introduction}
\lbl{sec:intro}
We study the asymptotic behaviour of a finite network of
oscillators (harmonic or anharmonic), where some of the oscillators are coupled to thermostats.
Our principle example is a chain of $N\geq 1$ oscillators interacting with two thermostats situated at the boundary of the chain.
One of the first rigorous results in this direction was obtained in \cite{EPRB}, where
the just mentioned case of chain was considered. The authors modelled the thermostats by the linear wave equations and
the interaction between the thermostats and the chain was chosen to be linear as well. The initial conditions of the thermostats were assumed to be random and distributed according to the Gibbs measures of given temperatures $\TT_L$ and $\TT_R$. 
Under appropriate assumptions the mixing property was established, stating that the asymptotic behaviour of the chain is governed by a stationary measure $\mu$, which is unique
and absolutely continuous with respect to the Lebesgue measure. 
More precisely, the authors proved the total variation convergence of measures
\bee\lbl{0conv}
\DD(p,q)(t)\to \mu \ass t\to\infty,
\eee
where $\DD\xi$ denotes the distribution of a random variable $\xi$ and $(p,q)(t)$ is the vector of coordinates and moments of the oscillators in the chain at the time $t$.
Moreover, in \cite{EPRBa} the authors proved that the stationary measure $\mu$ has positive entropy production.
These results were subsequently developed in \cite{RBT}, \cite{Car}-\cite{CPo}.

Because of the Gibbs distribution of initial conditions, 
 the initial energies $\EE_L(0),\EE_R(0)$ of the thermostats in the model above are almost surely infinite. 
 In the present work we address the question what happens if the initial conditions are chosen in such a way that the energies $\EE_L(0),\EE_R(0)$ are finite (but, probably, very large)? 
Namely, we consider a system similar to that investigated in \cite{EPRB}
but assume the initial conditions to be deterministic and 
the total energy of the system to be finite.
Under appropriate assumptions we show 
that when $t\to\pm\infty$ the oscillators in the chain just stop.
That is, we prove the convergence 
\bee\lbl{mainconv}
(p,q)(t)\to (0,q_c^{\pm}) 
\qnd
\fr{d^i}{dt^i} (p,q)(t)\ra 0
\qu \forall i\geq 1 
\ass t\to\pm\infty,
\eee
where the constants $q_c^{\pm}$  
are critical points of some effective potential.
The critical points $q_c^{\pm}$ may depend on the initial conditions,
however, the effective potential is independent of them, 
so that the set of all critical points $\{q_c^{\pm}\}$ does not depend on the initial conditions.

We next study the asymptotic behaviour of the energies $\EE_L$ and $\EE_R$ of the thermostats when $t\to\pm\infty$. 
In the case when the chain consists of a unique oscillator ($N=1$), 
we prove that the energies $\EE_L$ and $\EE_R$ 
converge to some constants $\EE_L^{\pm}$ and $\EE_R^{\pm}$ correspondingly, 
so that the energy transport between the thermostats 
stops as well.
Then, one could expect that $\EE_L^{\pm}=\EE_R^{\pm}$,
so that the system reaches a kind of thermal equilibrium.
We show, however, that this situation is not generic, but takes place only for a special set of initial conditions, 
of codimension one. 
Thus, 
the chain plays a role of an insulator rather than of a conductor.
For longer chains, when $N>1$, we can prove only 
that the sum $\EE_L+\EE_R$  and the time derivatives 
$\ds{\fr{d^i}{dt^i}\EE_L,\fr{d^i}{dt^i}\EE_R}$ converge as $t\to\pm\infty$, 
for any $i\geq 1$.
We conjecture, however, that the same results as for $N=1$ hold, 
probably under stronger assumptions for the function specifying the interaction between the thermostats and the chain,
and for the initial conditions of the thermostats.

We use a method 
developed by D.Treschev in \cite{Tr}. 
The main idea behind is that the motion of the oscillators in the chain cannot create parametric resonances in the thermostats: 
otherwise the total energy of the system could not be finite.
If the coupling of the thermostats with the chain is "sufficiently strong",
this implies a serious restriction for the dynamics of the chain which leads to 
the convergence \eqref{mainconv}.

The convergence \eqref{mainconv}
generalizes results obtained in 
\cite{Tr},\cite{Dym12} and \cite{Sau},
while the asymptotic behaviour of the energies $\EE_{L,R}$
is studied for the first time. 
In \cite{Tr}, under appropriate assumptions, the convergence (\ref{mainconv}) was established for the system of one oscillator ($N=1$) interacting with one thermostat.
In \cite{Dym12}, the result of \cite{Tr} was generalized for arbitrary network of $N$ oscillators
(not necessarily forming a chain)  interacting with one thermostat,
under the assumption that the oscillators are harmonic.
In \cite{Sau},  
the assumption of harmonicity was removed but instead it was assumed that 
the number $M$ of the thermostats interacting with the system of oscillators is not less 
then the number of oscillators, that is
\bee\lbl{M>N}
M\geq N.
\eee 
To establish the convergence (\ref{mainconv}), 
in the present paper we
 significantly relax the condition (\ref{M>N}) taking into account 
geometry of the network.
\footnote{For simplicity, we restrict ourselves to the case when each thermostat is allowed to interact with a unique oscillator. In contrast, in \cite{Sau} each thermostat is allowed to interact with several oscillators.} 
Namely, we prove  \eqref{mainconv}  under the assumption that the oscillators interacting with the thermostats 
"control" in appropriate sense the other oscillators from the network (see assumption A5 in Section~\ref{sec:ass}).
For example, our results apply to a chain of oscillators interacting through the boundary with at least one thermostat, as well as to a tree of oscillators, where the thermostats are coupled to the oscillators in the leafs of the tree. See Fig.\ref{fig2} for more examples. See also
Section~\ref{sec:outline} for comparison  of our strategy  with that used in \cite{Sau}.

When we were editing our manuscript, the paper \cite{CEHRB} was published,  where
the convergence \eqref{0conv} was generalized to the case of networks under exactly the same assumption A5,
which is called there the  controllability assumption. 
 
Let us note that for systems of the type \cite{EPRB}, where the energy is infinite and the coupling with the thermostats provides stochastic perturbation, an analogue of assumption \eqref{M>N} is well-known. 
Namely, one assumes that each oscillator from the network is coupled with its own thermostat.
This assumption significantly simplifies 
the investigation of the asymptotic dynamics and of the energy transport.
Indeed, under this assumption, the mixing property \eqref{0conv} is well understood  (see e.g. \cite{Ver87,Kha})
while concerning the energy transport there is a number of resent developments 
(see \cite{BLL,BLLO,BeKLeLu,Dym15,Dym16,Dym16a}).
Relaxation of assumption \eqref{M>N} in this setting is an important and complicated problem, and paper \cite{EPRB} provides one of the first results in this direction.

Effects similar to the convergence \eqref{mainconv} are also known in different infinite-dimensional Hamiltonian systems with finite total energy, mostly in the context of non-linear Hamiltonian PDEs. See a review \cite{Ko_r} and references therein. 
In particular, using a completely different method, in a similar but different setting in \cite{KSK} it was proven that a single oscillator interacting with the thermostat enjoys the convergence \eqref{mainconv}. 

The structure of the paper is as follows. 
In Section~\ref{sec:suar} we describe the model, state our
main results and give an outline of their proof. 
In Section~\ref{sec:prel} we establish some technical lemmas in spirit of \cite{Tr,Dym12,Sau}, playing a central role in the sequel.
In Section~\ref{sec:proofs} we prove our main results.
In Section~\ref{sec:oo} we study the asymptotic behaviour of the energies of two thermostats which are coupled to a unique oscillator.

\section{Setup and results}
\lbl{sec:suar}

\subsection{Setup}
\lbl{sec:setup}
Let us take a finite undirected graph $\Ga=(\GG, \mathfrak{E})$, where $\GG$ and $\mathfrak{E}$ stand for the sets of vertices and edges of $\Ga$ correspondingly. 
We consider a system of one-dimensional oscillators enumerated by the vertices $j\in\GG$ of the graph $\Ga$.  If  the vertices $i,j\in\GG$ are adjacent (we write $i\volna j$ or $(i,j)\in\mathfrak{E}$), we couple the corresponding oscillators by an interaction potential $V_{ij}$. The Lagrangian of the system has the form
\bee\non
\LL^O(q,\dot q)
	=\sum_{j\in\GG} \Big(\fr{\dot q_j^2}{2}-U_j(q_j)\Big) - \frac12\sum_{i,j\in\GG:\,i\volna j} V_{ij}(q_i-q_{j}),
\eee
where the dot stands for the derivative in time, 
$(q,\dot q)=(q_j,\dot q_j)_{j\in\GG}\in\mR^{2|\GG|}$,
$U_j,V_{ij}$ are smooth real functions and $V_{ij}(q_i-q_{j})=V_{ji}(q_j-q_{i})$.
We fix a set 
$$\La\subset\GG$$ 
and couple each oscillator from $\La$ with its own thermostat. 
Each thermostat is modelled as a continuum collection of independent harmonic oscillators parametrized
by their internal frequency $\nu$. 
The thermostats are given by the Lagrangians
\bee\lbl{Lagr}
\LL^T_m(\xi_m,\dot \xi_m)=\fr12\int_{-\infty}^\infty \dot \xi_m^2(\nu)-\nu^2\xi_m^2(\nu)\,d\nu,
\qquad m\in\La,
\eee 
where $\xi_m(\nu,t),\dot\xi_m(\nu,t)\in\mR$ stand for
the coordinate and velocity of the oscillator which has the internal frequency $\nu$.
\br
It is more natural to assume that the Lagrangians have the form
$$
\wt\LL^T_m(\zeta_m,\dot \zeta_m)=\fr12\int_{-\infty}^\infty \rho_m(\nu)\big(\dot \zeta_m^2(\nu)-\nu^2\zeta_m^2(\nu)\big)\,d\nu,
$$
where the physical meaning of the function $\rho_m(\nu)>0$ is the density of oscillators with the internal frequency $\nu$.
We consider the simplified Lagrangian \eqref{Lagr} 
since it can be obtained from the Lagrangian
$\wt\LL^T_m$ by the transformation 
$\xi_m=\sqrt{\rho_m}\zeta_m$, $\dot\xi_m=\sqrt{\rho_m}\dot\zeta_m$.
\er
The coupling between the system of oscillators and the thermostats is linear and given by the potentials 
\bee
\lbl{intpot}
V_m^{\intt}(q_m,\xi_m)=-q_m\int_{-\infty}^\infty \kappa_m(\nu)\xi_m(\nu)\, d\nu, 
\qquad m\in\La,
\eee
where $\ka_m$ are real continuous functions.
The Lagrangian of the total system has the form 
$$
\LL(q,\xi,\dot q,\dot\xi)
		=\LL^O(q,\dot q)+\sum_{m\in\La}\LL^T_m(\xi_m,\dot \xi_m)
		-\sum_{m\in\La} V^{\intt}_m(q_m,\xi_m),
$$  
where $(\xi,\dot\xi)=(\xi_m,\dot\xi_m)_{m\in\La}$.
Set $\de_{j\La}=1$ if $j\in\La$ and $\de_{j\La}=0$ otherwise.
Then the equations of motion take the form
\begin{align}\lbl{eq}
\ddot q_j&=-U_j'(q_j)+\sli_{i\in\GG:\,i\volna j}V_{ij}'(q_{i}-q_j) 
+ \de_{j\La}\int_{-\infty}^\infty\ka_j(\nu)\xi_j(\nu)\, d\nu, 
\\\lbl{eq2}
\ddot\xi_m(\nu)&=-\nu^2\xi_m(\nu)+\kappa_m(\nu)q_m, 
\qquad m\in\La,\, \nu\in\mR,\, j\in\GG,
\end{align}
where 
$U_j'(x)$, $V_{ij}'(x)$ denote the derivatives
of the functions $U_j,V_{ij}$ in $x$.
We fix some initial conditions
\bee\lbl{ini}
(q,\xi,\dot q,\dot\xi)(0)=(q_0,\xi_0,\dot q_0,\dot\xi_0).
\eee
The total energy of the system has the form
\begin{align}\non
E(q,\xi,\dot q,\dot\xi)
		&=\sum_{j\in\GG}\Big(\fr{\dot q_j^2}{2} +  U_j(q_j) \Big) + \frac12\sum_{i\volna j} V_{ij}(q_i-q_j) 
		+  \fr12\sum_{m\in\La}\intf \dot\xi^2_m + \nu^2\xi_m^2\,d\nu 
		\\ \lbl{energy0}
		&-\sum_{m\in\La} q_m\int_{-\infty}^\infty \kappa_m\xi_m\, d\nu.
\end{align}
\begin{rem}
In the paper \cite{EPRB} the authors considered 
a system similar to \eqref{eq}-\eqref{eq2}, 
where the graph $\Ga$ was chosen as a chain $\{1,\ldots,N\}$ of the length $N\geq 1$, and $\La=\{1,N\}$. The dynamics of the thermostats was given by the
wave equations 
$\partial_{tt}\ph_{m}(x,t)=\Delta\ph_m(x,t) + \al_m(x) q_m(t)$,
where $x\in\mR^d$.
If $d=1$, under the Fourier transform the wave equations take 
the form 
$\ddot {\hat\ph}_m(\nu,t)=-\nu^2\hat\ph_m(\nu,t) + \hat\al_m(\nu)q_m(t)$,
and the total system takes the form \eqref{eq}-\eqref{eq2}.
Note, however, that the effective potential arising in the formula (3.7) of \cite{EPRB} is different from our effective potential \eqref{effpot}, introduced in the next subsection.
\end{rem}

\subsection{Assumptions}
\lbl{sec:ass}
We impose on the system the following assumptions. 
\begin{description}
\item[A1] 
{\it The potentials $U_j,V_{ij}$ are smooth for every $i,j\in\GG$.
The second derivatives $V_{ij}''$ have only isolated zeros.}

\end{description} 
Denote by $L^n$ the space of measurable functions $f:\mR\mapsto\mC$ satisfying
$\ilif |f(x)|^n\, dx<\infty$. 
\begin{description}
\item[A2]
{\it The functions $\ka_m$ are $C^2$-smooth
and
$\ka_m(\nu)=0$ if and only if $\nu=0$.
Moreover, 
$\ka_m,\ka_m''\in L^1\cap L^2$ and there
exists an integer $r\geq 0$ such that
$\nu^r \ka,\, \nu^{r/2}\ka'\in L^2$.}
\end{description}
Denote
\bee\lbl{Km}
K_m:=\int_{-\infty}^\infty \fr{\ka_m^2}{\nu^2} \,d\nu<\infty,
\eee
where the inequality 
$K_m<\infty$ follows from
assumption A2.
Introduce the {\it effective potential} 
\bee\lbl{effpot}
V^{\eff}(q):=\sum_{j\in\GG} U_j(q_j) + \frac12\sum_{i\volna j}V_{ij}(q_i-q_{j})
- \sum_{m\in\La}\fr{K_m q_m^2}{2}.
\eee
\begin{description}
\item[A3]
{\it The effective potential 
$V^{\eff}$
satisfies
$\big|V^{\eff}(q)\big|\to \infty$ as $|q|\to\infty$.}
\end{description}
The total energy \eqref{energy0} of the system can be written in the form
\bee\lbl{energy}
E(q,\xi,\dot q,\dot\xi)
		=\sum_{j\in\GG}\fr{\dot q_j^2}{2} 
		+  \sum_{m\in\La}\intf\fr{\dot\xi^2_m}{2}\,d\nu + V^{\eff}(q)
		+ \sum_{m\in\La}\intf \fr{\nu^2}{2}\Big(\xi_m-\fr{\ka_m q_m}{\nu^2}\Big)^2\,d\nu.
\eee
Due to assumption A3, without loss of generality we can assume $V^{\eff}\geq 0$,
so that $E\geq 0$.
Denote by $\EE_m$ the energy of the $m$-th thermostat,
\bee\lbl{EEm}
\EE_m(\xi,\dot\xi):=
\fr12\intf\dot\xi_m^2(\nu)+ \nu^2\xi_m^2(\nu)\,  d\nu.
\eee
\begin{description}
\item[A4]
{\it The initial conditions \eqref{ini} satisfy
$\EE_m(\xi_0,\dot\xi_0)<\infty$ 
for any $m\in\La$.}
\end{description}
Assumption A4 together with \eqref{Km} implies 
that the initial total 
energy $E(q_0,\xi_0,\dot q_0,\dot\xi_0)$
is finite. 
Indeed, by the Cauchy-Bunyakovsky inequality we have 
$$
\ilif \ka_m\xi_{0m}\,d\nu\leq \Big(\ilif\fr{\ka_m^2}{\nu^2}\,d\nu\Big)^{1/2}
    \Big(\ilif\nu^2\xi_{0m}^2\,d\nu\Big)^{1/2}<\infty,
$$
so that, due to \eqref{energy0}, we have 
$E(q_0,\xi_0,\dot q_0,\dot\xi_0)<\infty$.

To formulate the next assumption, we construct a set $\La_\Ga$ by the following inductive procedure. 
We start by setting $\La_\Ga:=\La$ and consider a set $\La_\Ga^1$ of vertices $j\in\La_\Ga$ for which there exists a {\it unique} vertex $n(j)\in\GG\sm\La_\Ga$ adjacent to $j$ (in particular, for all other $i\in\GG$, $i\volna j$, we have $i\in\La_\Ga$).  We add $n(j)$ to $\La_\Ga$,
so that $\La_\Ga:=\cup_{j\in\La_\Ga^1}n(j)\cup\La_\Ga,$ and iterate the procedure. We finish when we get $\La_\Ga^1=\emptyset$. 
\begin{description}
\item[A5]
{\it We have $\La_\Ga=\GG$.}
\end{description}
See Fig.\ref{fig2} for several examples of networks which satisfy assumption A5. 
\begin{figure}[t]
	\centering
	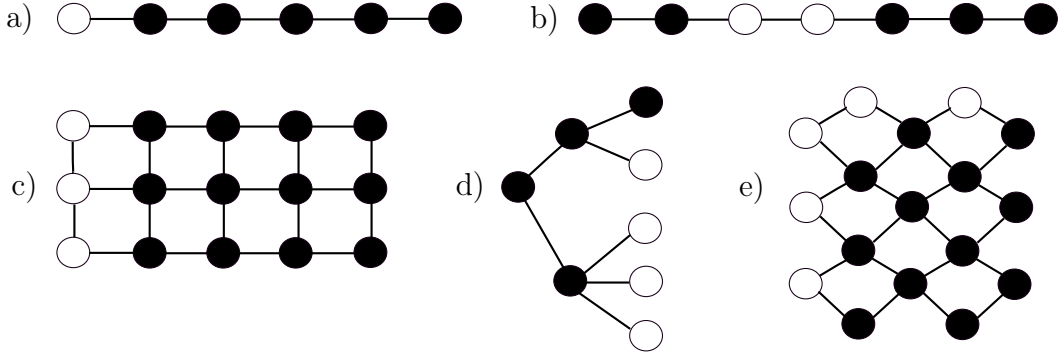
	\caption{Networks a)-e) satisfy assumption A5 provided that the set $\La$ contains the white oscillators. }
	\lbl{fig2}
\end{figure}

\begin{rem}\lbl{rem:A2}
	We give assumption A2 in the form above 
	for simplicity of formulation, while 
	in reality we use the following weaker assumption:
	
	The functions $\ka_m$ are continuous and 
	$\ka_m(\nu)=0$ if and only if $\nu=0$. Moreover,
	\bee\lbl{A2}
	\nu^l\ka_m\in L^2 \qnd \ka_m,\, \nu^l\ka_m^2\in\hat L^1,
	\qmb{for any}\qu -1\leq l\leq r,
	\eee
	where $\hat L^1=\FF(L^1)$ stands for the Fourier transform of the space $L^1$.
	It is straightforward to check that assumption A2 implies the relation \eqref{A2}. To do this, one should use the fact that a $C^2$-smooth function $g$ satisfying the inclusion $g, g'' \in L^1$, satisfies $\hat g\in L^1$, or, equivalently, $g\in\hat L^1$, 
	where $\hat g$ is the Fourier transform of the function $g$. Indeed, this follows from the relations 
	$
	\big|\hat g(\la)\big|\leq \|g\|_{L^1}
	$
	and 
	$ 
	\big|\la^2\hat g(\la)\big|
	\leq \|g''\|_{L^1},
	$ 
	which hold for any $\la$.
\end{rem}
\subsection{Main results}

Our main goal is to study the large-time asymptotic behaviour of the system of oscillators. 
However, since the total system has infinite dimension,
even its well-posedness is not immediate.
\bt\lbl{lem:existence}\emph{(\cite{Sau})}
Assume \emph{A1}-\emph{A4}. 
Then the system \eqref{eq}-\eqref{ini} has a unique
solution $q(t), \xi(\nu,t)$ 
and this solution is defined for all $t\in\mR$.
The energy 
computed on this solution is finite and does not depend on time,
$E(q,\xi,\dot q,\dot\xi)(t)\equiv E(q_0,\xi_0,\dot q_0,\dot\xi_0)<\infty$.
The functions $q_k(t)$, 
$\EE_j(q,\xi,\dot q,\dot\xi)(t)$ are uniformly bounded and the functions $q_k(t)$
are uniformly Lipschitz.
\et
Theorem \ref{lem:existence} is a particular case of \rtheo 3.1 from \cite{Sau};
see also  \rtheo 1 from \cite{Tr} for an earlier similar result.
The uniform Lipschitz property of the functions $q_k$ is not stated in the cited works,
but it follows immediately from the uniform boundedness of the derivatives $\dot q_k$
which takes place because the energy \eqref{energy} is constant.

Next we state our main results.
Let $\NN^{\eff}$ be a set of critical points of the effective potential, 
$$
\NN^{\eff}=\{q\in\mR^N:\, \p_{q_j} V^{\eff}(q)=0 \; \forall j\in\GG\}.
$$
The set $\NN^{\eff}$ is closed. 
Denote by $\dist(q,\NN^{\eff})$ the Euclidean distance in $\mR^{|\GG|}$ from the point $q$ 
to the set $\NN^{\eff}$.   
Recall that the integer $r$ is defined in assumption A2.
\bt\lbl{th:final}
Under assumptions \emph{A1}-\emph{A5}, 
the solution $q(t),\xi(\nu,t)$ of  equation \eqref{eq}-\eqref{ini} satisfies
\bee\lbl{main}
\dist\big(q(t), \NN_\pm^{\eff}\big)\ra 0 \ass t\to\pm\infty,
\eee
where $\NN^{\eff}_+,\NN^{\eff}_-$ are some connected components of the set $\NN^{\eff}$.
Moreover, the solution $q(t)$ is $C^{r+3}$-smooth and for its time derivatives $q^{(l)}$, 
$1\leq l\leq r+3$, we have
\bee\lbl{dertozero}
q^{(l)}(t)\ra 0 \ass t\to\pm\infty.
\eee
\et
The proof of Theorem \ref{th:final} is given in Section \ref{sec:final}.
The choice of the connected components $\NN^{\eff}_+,\NN^{\eff}_-\subset\NN^{\eff}$  may depend on the initial conditions \eqref{ini}.
Let us now additionally assume
\begin{description}
	\item[A6]
	{\it  The effective potential $V^{\eff}$ has only isolated critical points.}
\end{description}
Then any connected component of the set $\NN^{\eff}$ is a singleton. Consequently, Theorem~\ref{th:final} implies the convergence 
\bee\lbl{q_c}
q(t)\ra q_c^\pm \ass t\to\pm\infty,
\eee
where the critical points $q_c^+,q_c^-\in\NN^{\eff}$
may depend on the initial conditions. 
If $q_c^-\neq q_c^+$,
we observe a transition from one equilibrium at $t=-\infty$ 
to another one at $t=+\infty$.

As a corollary of Theorem~\ref{th:final} we obtain the following result
specifying the limiting as $t\to\pm\infty$ 
behaviour of the energies $\EE_m$ of the thermostats.

\begin{cor}\lbl{lem:entr}
	Under assumptions A1-A6,
	the energies of the  thermostats 
	$\EE_m(t)$, $m\in\La$,
	computed on the solution of equation \eqref{eq}-\eqref{ini}
	satisfy
	\bee\lbl{sumen}
	\sli_{m\in\La}\EE_m(t)\to E(q_0,\xi_0,\dot q_0,\dot\xi_0) - V^{\eff}(q_c^\pm)
	+ \sli_{m\in\La}\fr{K_m(q_c^\pm)^2}{2}
	\ass t\to\pm\infty,
	\eee
	where $q^\pm_c$ are the critical points of the effective potential given by \eqref{q_c}. 
	Moreover, the functions $\EE_m$ are $C^{r+1}$-smooth and their derivatives satisfy
	\bee\lbl{deren}
	\fr{d^l}{dt^l}\EE_m(t)\to 0 \ass t\to\pm\infty, 
	\qu \forall 1\leq l\leq r+1.
	\eee
\end{cor}
Corollary \ref{lem:entr} is established in Section \ref{sec:entr} and
it provokes the following questions. Is it true that
the energies $\EE_m(t)$ of the thermostats converge when $t\to\pm\infty$?
If this is the case, let us assume that the graph $\GG$ forms a chain $\{1,\ldots,N\}$ and
only the first and the last oscillators are coupled to the 
thermostats, that is $\La=\{1,N\}$.
Assume also that the functions $U_j,V_{ij},\ka_j$ are independent of 
$i,j$. 
Is it then true that the limits of 
the energies of the thermostats are equal, so that a kind of thermal equilibrium is achieved? 
We are able to give an answer to these questions only in a particular case.
Namely, in \rsec\ref{sec:oo} we consider a system of one oscillator interacting with two thermostats
and prove
that the energies of the thermostats converge. 
We show, however, that the set of initial conditions for which the limits coincide 
is of codimension one, so this situation is not generic.

Although we do not have a proof,
we believe that the same situation takes place
for longer chains as well, at least in the case when the functions $\ka_m$ together with the initial conditions $\xi_0, \dot\xi_0$ decay sufficiently fast at infinity.
In the case when the chain consists of a unique oscillator, we are able to establish the convergence of the energies because of the existence of a special change of variables (see \eqref{trans}). 
The latter transforms our system to a system consisting of one oscillator interacting with one thermostat, and of another isolated thermostat, so that the convergence of energies immediately follows from Corollary~\ref{lem:entr}. 
In the general case, however, an analogous transformation does not exist
and
the main ingredient we lack to prove the convergence of the energies $\EE_m$   is an appropriate estimate for the rate of convergence \eqref{main}. 

\subsection{Outline of the proof of Theorem \ref{th:final}}
\lbl{sec:outline}

The proof of the theorem relies on a method invented by D. Treschev in \cite{Tr} 
and subsequently developed in \cite{Dym12, Sau}.
To explain its main idea let us study equation \eqref{eq2}
where the function $q_m$ is viewed as an external force. 
Assume that $q_m(t)=\sin (\la t)$ or $q_m(t)=\cos (\la t)$ 
for some $\la\neq 0$. Then in equation  \eqref{eq2} with $\nu=\la$
the parametric resonance will occur,
since $\ka_m(\la)\neq 0$ by assumption A2. 
However, this contradicts to the energy conservation, 
so that such situation is impossible. 
This suggests that the function $q_m$ cannot have 
components oscillating with any frequency $\la\neq 0$.

To make these considerations rigorous, following \cite{Tr}
we introduce the notion of singular support  
$\ssupp \hat x$
of a distribution $\hat x$,
where $\hat x=\FF(x)$ is 
the Fourier 
transform of a bounded uniformly continuous real function $x$.
We say that $\la\in\ssupp\hat x$ if, roughly speaking,
the function $x$ has a component oscillating with the frequency $\la$;
see Definition~\ref{def:ssupp}. 
In particular, if $\ssupp\hat x=\emptyset$, we have 
$x(t)\ra 0$ as $t\ra\pm\infty$. 
The conservation  of energy together with the relation 
$\ka_m(\nu)\neq 0$ for $\nu\neq 0$ implies that
$\ssupp\hat q_m\subset\{0\}$ for $m\in\La$,
and it follows that $\ssupp\big(\la^l\hat q_m(\la)\big)=\emptyset$
for any $l\geq 1$,
since $\la^l\hat q_m(\la)$ vanishes at $\la=0$.
Since $\FF\big(q_m^{(l)}\big)(\la)=(i\la)^l\hat q_m(\la)$,
we obtain 
$\ssupp \FF\big(q_m^{(l)}\big)=\emptyset$,
so that
\bee\lbl{dertozerointro}
q_m^{(l)}(t)\ra 0 \ass t\to\pm\infty,
\qmb{for any}\qu
l\geq 1
\eee
and
$m\in\La$.
We then use the Duhamel formula to express the function
$
\ph_m(t):=\ilif \ka_m(\nu)\xi_m(\nu,t)\, d\nu
$
from equation \eqref{eq2}
through the function $q_m(t)$.
Using that $\ssupp\hat q_m\subset\{0\}$, we prove that
\bee\lbl{phintro}
\ph_m(t)=K_m q_m(t) + \thet_m(t),
\eee
where the constant $K_m$ is given by \eqref{Km}
and the function $\thet_m(t)\ra 0$ as $t\to\pm\infty$.

Assume first that $\La=\GG$,
so that each oscillator from the network is coupled with a thermostat
and, consequently, \eqref{dertozerointro} holds for all $m\in\GG$.
Then, inserting \eqref{phintro} into equation \eqref{eq},
letting $t\to\pm\infty$ and using $\eqref{dertozerointro}$
we obtain the desired convergence \eqref{main};
similar argument was used in the papers 
\cite{Tr} and \cite{Sau}.
In difference with these works, 
we do not have the relation
 $\La=\GG$,
so that an additional argument is needed
to establish \eqref{dertozerointro} for $m\notin\La$.
By construction of the set $\La_\Ga$,
due to assumption A5, 
there exists $j\in\La$ for which there is exactly one $n\in\GG\sm\La$ adjacent to $j$. In particular, for any $i\in\GG$ adjacent to $j$, $i\neq n$, we have $i\in\La$, so for such $i$ the convergence 
\eqref{dertozerointro} takes place.
Then we  differentiate in time the both sides of the $j$-th equation  from \eqref{eq} 
and find that all the terms of the resulting equation
vanish as $t\to\pm\infty$, except the term $V_{nj}''(q_n-q_j)\dot q_n$.
Thus, the latter term vanishes as well
and we prove that this implies the convergence
$\dot q_n(t) \ra 0$ as $t\to\pm\infty$. 
Using the uniform continuity of $\dot q_n$, 
we then establish the convergence \eqref{dertozerointro} for the $m=n$. 
Then we replace the set $\La$ by $\La\cup\{n\}$ and repeat the procedure. 
Finally, we get  \eqref{dertozerointro}  for all $m\in\La_\Ga$, that is 
for all $m\in\GG$ since, due to assumption A5, $\La_\Ga=\GG$. 
Arguing as above we conclude that this implies 
the convergence \nolinebreak \eqref{main}.

To follow the outlined above strategy we also need to develop some technical tools used in the papers
\cite{Tr,Dym12,Sau}. 
This is done in the next section.

\section{Preliminary results}
\lbl{sec:prel}

In this section we develop the method used in the papers \cite{Tr,Dym12,Sau} and obtain technical results playing a key role in the proof of \rtheo\ref{th:final}.

\subsection{Functional spaces}\lbl{sec:fs}

We denote by $\FF$ the Fourier transform and by $\FF^{-1}$ its inverse. 
We make a convention
\bee\lbl{Fourierr}
\hat\psi(\la)=\FF(\psi)(\la)=\ilif e^{-i\la t} \psi(t)\, dt
\qnd
\FF^{-1}(\hat\psi)(t)=\fr{1}{2\pi}\ilif e^{i\la t} \hat \psi(\la)\, d\la,
\eee
for $\psi\in L^1$. We also denote the Fourier transform by the "hat". Let us recall that for any $\ph,\psi\in L^1$ we have
$$
\FF(\ph\psi)=\fr{1}{2\pi}\hat\ph*\hat\psi \qnd \FF^{-1}(\hat\ph\hat\psi)=\ph*\psi,
$$
where $*$ stands for the convolution.

We denote by $\CC_b$ the space of {\it uniformly} continuous bounded functions $\psi:\,\mR\mapsto\mC$.
This is a Banach space with the standard norm 
$\|\psi\|_\infty=\sup\limits_{t\in\mR}|\psi(t)|$.
Let $\CC_0^+,\CC_0^-$ be subspaces of $\CC_b$
consisting of functions $\psi\in\CC_b$ satisfying 
$\psi(t)\ra 0$ as $t\ra\infty$ or $t\ra -\infty$ correspondingly.
Set $\CC_0:=\CC_0^+\cap\CC_0^-$, so that for $\psi\in\CC_0$
we have $\psi(t)\to 0$ as $t\to\pm\infty$.
We will also use the Fourier transforms $\hat\CC_b, \hat\CC^{\pm}_0$ and $\hat\CC_0$ of the spaces
$\CC_b, \CC^{\pm}_0$ and $\CC_0$, understood in the sense of distributions.

Let $\MM$ be the space of complex Radon measures of bounded variation on $\mR$.
This is a Banach space
with the norm
$$
\|\mu\|_{\MM} := \sup\limits_{\psi\in\CC_b:\,\|\psi\|_\infty=1} \lan \mu, \psi\ran,
\qmb{where}\quad
\lan \mu, \psi\ran:=\ilif \psi(\tau)\,\mu(d\tau).
$$
The Fourier transform $\hat\MM$ of the space $\MM$ is defined by
\bee\lbl{Fmeas}
\hat\mu(\la)=\ilif e^{-i\la t}\,\mu(d t),
\eee
and it is known that $\hat\MM\subset\CC_b$
(see e.g. \cite{Bog}). 
In particular,  we have $\hat L^1:=\FF(L^1)\subset \hat\MM$, 
since any measure $\mu\in\MM$ which is absolute continuous with respect to the Lebesgue measure
has the form $\mu(dt)=\tilde\mu(t)\,dt$ with $\tilde\mu\in L^1$.
By $\FF^{-1}(\hat\mu)$ we write the inverse Fourier transform of $\hat\mu\in\hat\MM$,
which is defined as the action of the operator $\FF^{-1}$ from \eqref{Fourierr} on the space $\CC_b$ and is understood in the sense of distributions. 
 Thus, $\FF^{-1}(\hat\mu)$ is a distribution $\tilde\mu$ which can be viewed as a density of the measure $\mu$, so that, informally, $\mu(dt)=\tilde\mu(t)\,dt$. 
We identify $\mu$ with $\tilde\mu$ and abusing notation
sometimes we write  $\mu(t)\,dt$ instead of $\mu(dt)$.
For $\hat\psi\in\hat\CC_b$ and $\hat\mu\in\hat\MM$
we define the pairing 
\bee\lbl{fourier_pairing}
\lan \hat\mu,\hat\psi\ran:=2\pi\lan \mu,\bar\psi\ran,
\eee
where $\psi=\FF^{-1}(\hat\psi)$ and
$\mu=\FF^{-1}(\hat\mu)$.  
In particular, when $\mu(d t)=\wt\mu(t)\,dt$ and the functions $\wt\mu,\psi$ belong to the Schwartz class, we have 
\footnote{Usually one defines the Fourier transform of measures via \eqref{Fmeas} with the exponent $e^{-i\la t}$ replaced by $e^{i\la t}$. The sign $"-"$ in the definition is more convenient for us since we want to have the same formula for the Fourier transform of spaces $\MM$ and $\CC_b$. The price we pay is the conjugation arising in the formula \eqref{fourier_pairing}.}
$\lan \hat\mu,\hat\psi\ran=\ili \hat\mu(\nu)\ov{\hat\psi(\nu)}\,d\nu
=2\pi \ili \mu(\nu)\ov{\psi(\nu)}\,d\nu$.
\begin{lem}\lbl{mu1mu2}
Let $\hat\mu,\hat\mu_1,\hat\mu_2\in\hat\MM$, $\hat\psi\in\hat\CC_b$ and $\hat\phi\in\hat\CC_0$. Then we have
$\hat\mu_1\hat\mu_2\in\hat\MM$,
$\hat\mu\hat\psi\in\hat\CC_b$
and 
$\hat\mu\hat\phi\in\hat\CC_0$.
\end{lem}
Proof of the lemma can be obtained via the Fourier transform in a standard way.
We do not present it but address the reader to \cite{Tr} (Section 8.1 and Lemma 8.2).
In this connection (and for further needs) we note that we have the following relations:
$$
\FF^{-1}(\hat\psi\hat\mu) 
= \psi*\mu=\ilif \psi(t-s)\,\mu(ds)
\qnd
\FF^{-1}(\hat\mu_1\hat\mu_2) 
= \mu_1*\mu_2,
$$
where the convolution $\mu_1*\mu_2\in\MM$ is defined by the formula 
$$
\lan\mu_1*\mu_2,\psi\ran=\ilif\ilif \psi(t+s)\,\mu_1(dt)\mu_2(ds), 
\qmb{where}\qu \psi\in\CC_b.
$$ 

For a more detailed discussion of the spaces $\CC_b,\CC_0,\MM$ and their Fourier transforms see~\cite{GT}. 
\bl\lbl{lem:m_dec}
Let $\hat\mu\in\hat\MM$, $a\in\mR$ and $\hat\mu(a)=0$. Then for any $\eps>0$ there exists $\hat\mu^\eps\in\hat\MM$ satisfying 
$\supp \hat\mu^\eps\subset (a-\eps,a+\eps)$ and
$a\notin\supp(\hat\mu-\hat\mu^\eps)$,
such that 
$\|\mu^\eps\|_{\MM}\to 0$ as $\eps\to 0$,
where $\mu^\eps:=\FF^{-1}(\hat\mu^\eps)$.
\el
{\it Proof.}
Without loss of generality we assume that $a=0$.
Let $\hat\chi$ be a smooth real function satisfying $\supp\hat\chi\subset [-1,1]$ and $\hat\chi(\nu)=1$ 
for any $|\nu|\leq 1/2$.
Set 
$$
\hat\chi^\eps(\nu):=\hat\chi(\nu/\eps) \qnd \hat\mu^\eps:=\hat\chi^\eps\hat\mu.
$$
Then it remains to show that $\|\mu^\eps\|_{\MM}\ra 0$ as $\eps\ra 0.$
Since 
$\mu^\eps=\mu*\chi^\eps$,
where
$\chi^\eps=\FF^{-1}(\hat\chi^\eps)$,
we have 
$$
\|\mu^\eps\|_{\MM}=\sup\limits_{\ph\in\CC_b:\,\|\ph\|_\infty=1}
\big|\lan\mu*\chi^\eps,\ph\ran\big|
=\sup\limits_{\ph\in\CC_b:\,\|\ph\|_\infty=1} \big|\lan\mu,\ph*\wt\chi^\eps\ran\big|,
$$
where $\wt\chi^\eps(t):=\chi^\eps(-t)$.
Set $b:=(\ph*\wt\chi^\eps)(0)$.
Since $\hat\mu(0)=0$, we have 
$\lan\mu,b\ran=b\hat\mu(0) = 0$. 
Then,
denoting 
by $\mI_{A}$ the indicator function 
of a set $A$,
we find
\begin{align*}
\lan\mu,\ph*\wt\chi^\eps\ran&=
\lan\mu,\ph*\wt\chi^\eps-b\ran
=\big\lan\mu,\mI_{\{|t|\leq \eps^{-1/2}\}}\big(\ph*\wt\chi^\eps-b\big)\big\ran
+\big\lan\mu,\mI_{\{|t|> \eps^{-1/2}\}}\big(\ph*\wt\chi^\eps-b\big)\big\ran
\\
&=:I_1^\eps+I_2^\eps.
\end{align*}
Let us estimate the term $I_1^\eps$.
Using that $\wt\chi^\eps(t)=\chi^\eps(-t)=\eps\chi(-\eps t)$, we obtain
\begin{align*}
|I_1^\eps| 
&\leq \|\mu\|_{\MM}\,
    \big\|\mI_{\{|t|\leq \eps^{-1/2}\}}\big(\ph*\wt\chi^\eps-b\big)\big\|_\infty
= \|\mu\|_{\MM} \sup\limits_{|t|\leq\eps^{-1/2}}
    \Big|\ilif \ph(y)(\wt\chi^\eps(t-y)-\wt\chi^\eps(-y))\,dy\Big| 
\\
&\leq \|\mu\|_{\MM}\sup\limits_{|t|\leq\eps^{-1/2}}
    \eps\ilif |\chi(\eps( y-t))-\chi(\eps y)|\, dy
=\|\mu\|_{\MM}\sup\limits_{|t|\leq\eps^{-1/2}}
    \ilif |\chi(y-\eps t)-\chi(y)|\, dy,
\end{align*}
so that $I_1^\eps\ra 0$ as $\eps\ra 0$,
uniformly in $\ph\in\CC_b$ satisfying $\|\ph\|_\infty=1$.
To estimate the term $I_2^\eps$, we recall that for any $\mu\in\MM$ there exist non-negative $\mu_1,\ldots,\mu_4\in\MM$ satisfying 
$\mu_1-\mu_2+i\mu_3-i\mu_4=\mu$, see e.g. \cite{Bog}.
Then,
$$
|I_2^\eps|\leq \sli_{j=1}^4 \lan\mu_j,\mI_{\{|t|> \eps^{-1/2}\}}\rangle
\|\ph*\wt\chi^\eps-b\|_\infty
\leq 2\|\ph*\wt\chi^\eps\|_\infty\sli_{j=1}^4 \lan\mu_j,\mI_{\{|t|> \eps^{-1/2}\}}\rangle.
$$
We have
$$
\|\ph*\wt\chi^\eps\|_\infty \leq \|\ph\|_\infty \|\wt\chi^\eps\|_{L^1}=\|\chi\|_{L^1}.
$$
Since for any $\al\in\MM$ we have
$\lan\al,\mI_{\{|t|> \eps^{-1/2}\}}\rangle\to 0$ as $\eps\to 0$,
we obtain
$I_2^\eps\ra 0$ as $\eps\to 0$,
uniformly in $\ph\in\CC_b$ satisfying $\|\ph\|_\infty=1$.
\qed

\subsection{Singular support}

Here we discuss the central for our paper notion of {\it singular support},
first introduced in \nolinebreak \cite{Tr}. 
\begin{defi}\lbl{def:ssupp}(\cite{Tr})
Let $\hat\phi\in\hat\CC_b$, $\la\in\mR$ and $\star\in\{+,-\}$. We say that $\la\in\ssupp^\star \hat\phi$ if for any interval $I\subset\mR$ containing the point $\la$ there exists $\hat\mu\in\hat\MM$, $\supp\hat\mu\subset I$, such that $\hat\mu\hat\phi\notin\hat\CC_0^\star$. 
We set
$
\ssupp\hat\phi:=\ssupp^+\hat\phi\cup\ssupp^-\hat\phi.
$
\end{defi}
For example, if $\hat\phi(\la) =\delta(\la-\la_0)$, where $\la_0\in\mR$ 
and $\delta$ is the Dirac delta-function, 
we have $\ssupp \hat\phi=\{\la_0\}$.
On the other hand, for $\hat\ph\in\hat\CC_0$ we have 
$\ssupp\hat\ph=\emptyset$
since by Lemma~\ref{mu1mu2} we have
$\hat\mu\hat\ph\in\hat\CC_0$, for any $\hat\mu\in\hat\MM$.
\bl\lbl{lem:ssupp} Let $\hat\phi\in\hat\CC_b$ and $\hat\mu\in\hat\MM$.
\begin{enumerate}
\item
Assume that  
$\hat\mu\hat\phi\in\hat\CC_0$. 
Then $\ssupp\hat\phi\subset\{\nu\in\mR:\,\hat\mu(\nu)=0\}$.
\item 
Assume that $\ssupp\hat\phi=\emptyset$
or 
$\ssupp\hat\phi\subset\{a_1,\ldots,a_n\}$
for $a_1,\ldots,a_n\in\mR$, $n\geq 1$,
and $\hat\mu(a_1)=\ldots=\hat\mu(a_n)=0$.
Then $\hat\mu\hat \phi\in\hat\CC_0$.
Moreover,
the equality $\ssupp\hat\phi=\emptyset$ implies $\hat\phi\in\hat\CC_0$.
\end{enumerate}
\el
{\it Proof.}
Item 1 is proven in  Lemma 8.9 of \cite{Tr}.
If $\ssupp\hat\phi=\emptyset$ then 
Lemma 8.6 from \cite{Tr} implies that $\hat\mu\hat\phi\in\hat\CC_0$. 
Choosing $\hat\mu=1$ we get $\hat\phi\in\hat\CC_0$.

It remains to study the case when 
$\ssupp\hat\phi\subset\{a_1,\ldots,a_n\}$.
We apply \rlem\ref{lem:m_dec} to $\hat\mu$
at the point $a_1$ and construct $\hat\mu^\eps_1\in\hat\MM$ 
satisfying $\|\mu^\eps_1\|_{\MM}\ra 0$ as $\eps\ra 0$,
$a_1\notin\supp(\hat\mu-\hat\mu^\eps_1)$
and 
$a_2,\ldots,a_n\notin\supp \hat\mu^\eps_1$, for $\eps$ 
sufficiently small.
Next we apply \rlem\ref{lem:m_dec} to $\hat\mu-\hat\mu^\eps_1$ 
at the point $a_2$ and obtain 
$\hat\mu^\eps_2\in\hat\MM$ 
satisfying $\|\mu^\eps_2\|_{\MM}\ra 0$ as $\eps\ra 0$,
$a_1,a_2 \notin\supp(\hat\mu-\hat\mu^\eps_1-\hat\mu^\eps_2)$
and $a_3,\ldots,a_n\notin\supp \hat\mu^\eps_2,$
for $\eps$ sufficiently small.
Iterating the procedure,
we construct $\hat\mu^\eps_3,\ldots,\hat\mu^\eps_n\in\hat\MM$ satisfying
\bee\lbl{muto0}
\|\mu^\eps_j\|_{\MM}\ra 0 \ass \eps\ra 0
\eee
and 
\bee\lbl{aaanotinmu}
a_1,\ldots,a_n\notin\supp(\hat\mu-\hat\mu^\eps_1-\ldots -\hat\mu^\eps_n).
\eee
Lemma 8.6 from \cite{Tr} states that the relation \eqref{aaanotinmu} together
with the assumption
$\ssupp\hat\phi\subset\{a_1,\ldots,a_n\}$ provides that
$(\hat\mu-\hat\mu^\eps_1-\ldots -\hat\mu^\eps_n)\hat\phi\in\hat\CC_0$.
On the other hand, in view of (\ref{muto0}), we have
$\|\FF^{-1}(\hat\mu^\eps_j\,\hat\phi)\|_\infty\leq \|\mu^\eps_j\|_{\MM}\|\phi\|_\infty\ra 0$ as $\eps\ra 0$.
This implies the desired inclusion $\hat\mu\hat\phi\in\hat\CC_0$.
\qed
\bl\lbl{lem:der}
Assume that a $C^l$-smooth, $l\geq 1$, real function $\phi$ satisfies
$\ssupp \hat \phi\subset\{0\}$ and its derivatives $\phi^{(k)}$, $0\leq k\leq l$, belong to the space $\CC_b$.
Then $\phi^{(k)}\in\CC_0$ for any $1\leq k\leq l$.
\el
In particular, the assumption $\ssupp \hat \phi\subset\{0\}$ is satisfied 
if $\phi\in\CC_0$, since in this case $\ssupp \hat \phi=\emptyset$.

{\it Proof.}
Note that
$\FF\big(\phi^{(k)}\big)(\nu)=(i\nu)^k\hat \phi(\nu)$.
We first claim that for any $\hat\mu\in\hat\MM$ with compact support 
we have
$(i\nu)^k\hat\mu \in\hat\MM$.
Indeed, take a smooth function $\hat\chi$ with compact support satisfying $\hat\chi(\nu)=1$ for all $\nu\in\supp\hat\mu$.
Then we have
$(i\nu)^k\hat\chi  \in \hat\MM$, 
since $(i\nu)^k\hat\chi$ is a Schwartz function and Schwartz functions belong to the space $\hat\MM$.
Then,  
$(i\nu)^k\hat\mu = ((i\nu)^k\hat\chi)\hat\mu \in\hat\MM$,
by \rlem\ref{mu1mu2}.

Since $(i\nu)^k\hat\mu(\nu)$ equals to zero at the point $\nu=0$,
\rlem \ref{lem:ssupp}(2) implies that $(i\nu)^k\hat\mu \hat \phi\in\hat\CC_0$.
Thus, by the definition of singular support, we find $\ssupp \big((i\nu)^k\hat \phi\big)=\emptyset$.
Then, applying once more \rlem \ref{lem:ssupp}(2), 
we find $(i\nu)^k\hat \phi\in\hat\CC_0$ and consequently $\phi^{(k)}\in\CC_0$.
\qed

\subsection{Main lemma}

In this section we establish the following proposition 
which plays a key role in the proof of \rtheo\ref{th:final}.
Let $(q(t),\xi(t))$ be the solution of equation \eqref{eq}-\eqref{ini}.
Set
$$
\ph_m(t):=\ilif\ka_m (\la)\xi_m(\la,t)\, d\la.
$$
Recall that the numbers $r$ and $K_m$ are defined in assumption A2 and \eqref{Km}.
\bp\lbl{th:ssupp}
Under assumptions \textit{A1-A4}, for any $m\in\La$ we have
\begin{enumerate}
\item
$\ssupp\hat q_m\subset\{0\};$
\item
the function $\ph_m$
is $C^{r+1}$-smooth and all its derivatives belong to the space $\CC_b$.
The functions $q_j$, $j\in\GG$, are $C^{r+3}$-smooth and
their derivatives belong to $\CC_b$ as well.
\item
$\ph_m - K_m q_m  =\thet_m,$ where
$\thet_m\in\CC_0$. The function $\thet_m$ is $C^{r+1}$-smooth and its derivatives belong to the space $\CC_0$.
\end{enumerate}
\ep
\rprop\ref{th:ssupp} is a corollary of a result 
obtained in \rtheo 3.2 of \cite{Sau},
which we formulate below, in Theorem \ref{lem:sau}.  
See also 
\rtheo 2 from \cite{Tr} 
and \rtheo 2.2 from \cite{Dym12}
for similar results.
Set 
$\ds{\hat w_m(\nu):=\fr{2\pi \ka_m^2(\nu)}{i\nu}}$ 
and $w_m:=\FF^{-1}(\hat w_m)$.
By \eqref{A2}, $\ka^2_m/\nu\in\hat L^1$, 
so that $\hat w_m\in\hat L^1$ and $w_m\in L^1$.
Denote
\bee\lbl{w_def}
w_{\diamond \, m}^\pm(\tau):=
\fr12 \big(\mI_{[0,\pm\infty)}(\tau)(w_m(\tau) - w_m(-\tau))\big),
\eee
where $\mI_{[0,-\infty)}:=-\mI_{(-\infty,0]}$.
Using the formula (see e.g. \cite{Vl}),
$$
\FF(\mI_{[0,\pm\infty)})(\la) = \pm\pi\de(\la)+\fr{1}{i} v.p. \fr{1}{\la},
$$ 
where $v.p.$ means the principal value, we find
\bee\lbl{wdiad}
\hat w_{\diamond \, m}^\pm(\nu)
=\FF(w_{\diamond \, m}^\pm)(\nu)
=\pm\fr{1}{4}\big(\hat w_m(\nu)-\hat w_m(-\nu)\big)
+\fr{1}{2\pi i}v.p. \ilif\fr{\la \hat w_m(\la)}{\nu^2-\la^2}\,d\la.
\eee
Since $w_m\in L^1$, we find $w_{\diamond \, m}^\pm \in L^1$
and $\hat w_{\diamond \, m}^\pm\in\hat L^1$.
\bt\lbl{lem:sau}\emph{(\cite{Sau})} 
Under assumptions A1-A4 for any $m\in\La$ we have
\begin{enumerate}
\item
$\hat w_m\hat q_m \in L^1\subset \hat\CC_0;$

\item
$\hat \ph_m - \hat w_{\diamond \, m}^\pm \hat q_m 
\in \hat\CC_0^\pm.$
\end{enumerate}
\et
Despite that Theorem~\ref{lem:sau} is proven in \cite{Sau},
for the sake of consistency as well as for the further need we give its proof below. 

{\it Proof of Theorem~\ref{lem:sau}.}
{\it Item 1.} Using the Duhamel formula, from equation \eqref{eq} we find
\bee\lbl{xi12}
\xi_m(\nu,t)=\xi_{m}^0(\nu,t)+\xi_{m}^1(\nu,t),
\eee
where 
$$
\xi_{m}^0(\nu,t):=\Re\Big(\fr{1}{i \nu}{\bm \xi}_{0m} e^{i\nu t}\Big), 
\qu {\bm \xi}_{0m}:=\dot \xi_{0m} + i\nu\xi_{0m},
$$
and
\bee\lbl{xi1a}
\xi_{m}^1(\nu,t): = \fr{\ka_m}{2i\nu}\ili_0^t \big(e^{i\nu (t - \tau)} - e^{i\nu (\tau - t)}\big) q_m(\tau) \,d\tau
= \Re\Big(\fr{\ka_m}{i\nu}\hat q^t_m e^{i\nu t}\Big),
\eee
where
$q_{m}^t:=q_m \mI_{[0,t]}$ if $t\geq 0$, 
$q_{m}^t:=-q_m \mI_{[t,0]}$ if $t< 0$, 
and $\hat q_{m}^t :=\FF(q_{m}^t)$.
By a direct computation, from equation \eqref{xi12} we obtain
\bee\lbl{ennorm}
\EE_m(t)=\|\ka_m\hat q_{m}^t + {\bm \xi_{0m}}\|_{L^2}^2.
\eee
By assumption A4, we have $\|{\bm \xi_{0m}}\|_{L^2}^2=\EE_m(0)<\infty$
and, by Theorem~\ref{lem:existence}, 
the function $\EE_m(t)$ is bounded.
Then 
\bee\lbl{normqt}
\|\ka_m\hat q_{m}^t\|_{L^2}< C,
\eee
where the constant $C$ does not depend on the time $t$.
Set 
$
q_{m}^\pm:=q_m \mI_{[0,\pm\infty)},
$
where
$
\mI_{[0,-\infty)}:= -\mI_{(-\infty,0]},
$
and let $\hat q_{m}^\pm :=\FF(q_{m}^\pm)$.
\begin{lem}\lbl{lem:q+}
We have $\ka_m\hat q_{m}^\pm\in L^2$ and the functions $\ka_m\hat q_{m}^t$ weakly converge in $L^2$ to 
$\ka_m\hat q_{m}^\pm$ as $t\to\pm\infty$.
\end{lem}
{\it Proof.}
We first claim that the functions $\ka_m\hat q_{m}^t$ considered as 
functionals on the space $\hat\MM$ 
converge  
to $\ka_m\hat q_{m}^+$ as $t\to +\infty$.
Indeed, due to \eqref{fourier_pairing}, for any $\hat \mu\in\hat\MM$ we have
$$
\lan \hat\mu , \ka_m\hat q_m^t\ran = 2\pi \lan \mu, \ov{\FF^{-1}(\ka_m\hat q_m^t)} \ran 
= 2\pi \lan \mu*\FF^{-1}(\ka_m), q_m^t \ran. 
$$
Since, by \eqref{A2}, we have $\ka_m\in\hat L^1\subset\hat\MM$, we obtain $\mu*\FF^{-1}(\ka_m)\in \MM$
(see Lemma~\ref{mu1mu2}).
Then it is straightforward to check that  
$2\pi \lan \mu*\FF^{-1}(\ka_m), q_m^t \ran$ 
converges to 
$$
 2\pi \lan  \mu*\FF^{-1}(\ka_m),  q_m^+\ran = \lan \hat\mu, \ka_m\hat q_m^+ \ran.
$$
Due to the obtained convergence, the norm of the functional $\ka_m\hat q_m^+$,
considered as a functional on $\hat\MM\cap L^2$, is bounded by the 
constant from \eqref{normqt}. 
Since  the space $\hat\MM\cap L^2$ is dense in $L^2$,
we can uniquely extend $\ka_m\hat q_m^+$ to a continuous functional  on $L^2$ with the same norm.
Identifying the obtained functional with an element of $L^2$, we get the desired result.
The case $t\to -\infty$ can be studied similarly.
\qed

Since $\hat q_m = \hat q_m^+ - \hat q_m^-$, Lemma \ref{lem:q+} implies the inclusion $\ka_m\hat q_m \in  L^2.$
Since, by \eqref{A2}, we have $\ka_m/\nu\in L^2$, we find
$\ds{\hat w_m \hat q_m=\fr{2\pi\ka_m}{i\nu}\ka_m\hat q_m \in L^1}$.

{\it Item 2.}
Due to \eqref{xi12}, we have
$\ph_m=\ph_{m}^0+\ph_{m}^1$, where
$\ph_{m}^j:= \ilif \ka_m\xi_{m}^j\,d\nu$, $j=0,1$. 
Then
\bee\lbl{ph0a}
\ph_{m}^0 = \Re \ilif \fr{\ka_m}{i\nu} {\bm\xi}_{0m} e^{i\nu t}\,d\nu
= \Re\FF^{-1}\Big(\fr{2\pi\ka_m}{i\nu}{\bm \xi}_{0m}\Big).
\eee
Since by \eqref{A2} and assumption A4 
we have $\ka_m/\nu, {\bm\xi_{0m}} \in L^2$, we obtain $\ds{\fr{2\pi\ka_m}{i\nu}{\bm \xi}_{0m}}\in L^1$,
so 
\bee\lbl{phh1}
\ph_{m}^0\in\CC_0.
\eee
Next,
\bee\lbl{ph1a}
\ph_{m}^1= \Re\ilif \fr{\ka_m^2}{i\nu}\hat q_m^{t} e^{i\nu t}\, d\nu=:\Re \AAA(t).
\eee
We have
\bee\lbl{ph1b}
\AAA(t)=\FF^{-1} \big(\hat w_m \hat q_m^t\big)(t)
=(w_m*q_m^{t})(t)=\AAA_1^\pm(t)-\AAA_0^\pm(t),
\eee
where 
$$
\AAA_1^\pm(t):=\ilif q_m(t-\tau)\mI_{[0,\pm\infty)}(\tau) w_m(\tau) \, d\tau
$$
and
$$
\AAA_0^+(t):=\ili_t^\infty q_m(t-\tau) w_m(\tau) \, d\tau,  
\quad
\AAA_0^-(t):=\ili_{-\infty}^t q_m(t-\tau) w_m(\tau) \, d\tau.
$$
Since $q_m$ is real and $\hat w_m$ is purely imaginary,
we find
$$
\Re \AAA_1^\pm = \fr12 \big(\mI_{[0,\pm\infty)}(\tau)(w_m(\tau) - w_m(-\tau))\big)*q_m=\FF^{-1}(\hat w_{\diamond m}^\pm \hat q_m). 
$$
Moreover, since $w_m \in L^1$ we have $\AAA^\pm_0(t)\to 0$ as $t\to\pm\infty$.  
Consequently,
$$
\ph_m^1=\Re\AAA^\pm_1(t)- \Re\AAA^\pm_0(t) = \FF^{-1}(\hat w_{\diamond m}^\pm \hat q_m) + \thet^{\pm}_m,
$$
where $\thet^\pm_m\in\CC_0^\pm$.
Thus, in view of \eqref{phh1}, we have
$
\ph_m=\FF^{-1}(\hat w_{\diamond m}^\pm \hat q_m) + \wt\thet^{\pm}_m,
$
where $\wt\thet^\pm_m\in\CC_0^\pm$.
\qed

\ssk
Now we deduce Proposition \ref{th:ssupp} from Theorem~\ref{lem:sau}.

{\it Proof of Proposition \ref{th:ssupp}.} 
{\it Item 1.} 
Due to Theorem \ref{lem:sau}(1), we have $\hat w_m\hat q_m \in L^1\subset\hat\CC_0$.
Then \rlem\ref{lem:ssupp}(1)  implies that
$\ssupp\hat q_m\subset\{\nu:\,\hat w_m(\nu)=0\}$. 
Since, due to assumption A2, $\{\nu:\,\hat w_m(\nu)=0\}=\{0\}$,
we obtain the desired result.

{\it Item 2.} 
The result follows in a standard way from equations of motion \eqref{eq}-\eqref{eq2}.
Namely, due to the formulas \eqref{ph0a}-\eqref{ph1b},
we have 
$\ph_m=\ph_m^0+\ph_m^1$,
where 
$\ph_m^0\in\CC_0\subset\CC_b$ and $\ph_m^1(t)=\Re (w_m*q_m^t)(t)$.
Using the inclusion $w_m\in L^1$ it can be checked that
$\ph_m^1\in\CC_b$, so that $\ph_m\in\CC_b$.

For the time derivatives of the function $\ph_m$ we have
$\ph_m^{(l)}=(\ph_m^0)^{(l)}+(\ph_m^1)^{(l)}$.
By \eqref{ph0a} we obtain
$(\ph_m^0)^{(l)}=\Re\FF^{-1}(2\pi (i\nu)^{l-1}\ka_m {\bm \xi}_{0m})$.
Since by assumptions A2 and A4 we have 
$\nu^{r}\ka_m,  {\bm \xi}_{0m} \in L^2$,
we find $(\ph_m^0)^{(l)}\in \FF^{-1}(L^1)\subset\CC_0\subset\CC_b$,
for any $l\leq r+1$.

Let us now study the derivatives $(\ph_m^1)^{(l)}$.
Due to \eqref{xi1a}, we have
$$
\dot\ph_m^1(t)=\Re \int_{-\infty}^\infty \ka_m^2(\nu) e^{i\nu t}\int_0^t e^{-i\nu\tau}q_m(\tau)\,d\tau d\nu= \Re\FF^{-1}(2\pi \ka_m^2 \hat q_m^t )(t)
=\Re\big(\FF^{-1}(\ka_m^2)* q_m^t\big)(t). 
$$
Since by \eqref{A2} we have $\FF^{-1}(\ka_m^2)\in L^1$, we find
$\dot \ph_m^1\in\CC_b$ and, consequently, $\dot \ph_m\in\CC_b$. 
Thus, the r.h.s. of equation \eqref{eq} 
is differentiable in time and its derivative is from the space $\CC_b$.
Then the l.h.s. also is, so that the functions $q_j$ are differentiable at least three times 
and $ \dddot q_j\in\CC_b$, for any $j\in\GG$.
Next we compute $\ddot\ph_m^1$. Arguing as above, we find
$$
\ddot\ph_m^1(t)=
\big(\FF^{-1}((i\nu)\ka_m^2)* q_m^t\big)(t) + q_m(t)\int_{-\infty}^\infty \ka_m^2(\nu)\, d\nu.
$$
Then, using that $\FF^{-1}((i\nu)\ka_m^2) \in L^1 $, we obtain $\ddot\ph_m^1\in\CC_b$, 
so that $\ddot\ph_m\in\CC_b$. 
Differentiating equation \eqref{eq} two times, we then find $ q_j^{(4)}\in\CC_b$. 
Continuing the procedure in the same way, we arrive at the desired result.

{\it Item 3.} 
Set $\hat\de^\pm_m(\nu):=\hat w_{\diamond \, m}^\pm(0)-\hat w_{\diamond \, m}^\pm(\nu)$.
Since by \eqref{wdiad} we have $\hat w_{\diamond \, m}^\pm(0)=K_m$,
Theorem \ref{lem:sau}(2) implies
$$
\hat \ph_m  - K_m \hat q_m + \hat\de^\pm_m \hat q_m\in \hat\CC_0^\pm.
$$
Since $\hat\de^\pm_m(0)=0$, \rlem \ref{lem:ssupp}(2) together with 
the first item of the present proposition implies that
$\hat\de^\pm_m\hat q_m\in\hat\CC_0$, so that 
$\thet_m:=\ph_m-K_mq_m\in\CC_0^+\cap\CC_0^-=\CC_0.$

Due to item 2 of the proposition, the functions $\thet_m$ are $C^{r+1}$-smooth 
and their derivatives belong to the space $\CC_b$. 
Since $\thet_m\in\CC_0$ we have $\ssupp \hat\thet_m = \emptyset$, so the fact that the derivatives of $\thet_m$ belong to the space $\CC_0$
follows from \rlem\ref{lem:der}.
\qed

\section{Proofs of main results} 
\lbl{sec:proofs}

\subsection{Proof of Theorem \ref{th:final}}
\lbl{sec:final}

{\it Step 1.}
\rprop\ref{th:ssupp}(3) implies that equation \eqref{eq} 
can be written in the form
\bee\lbl{eq1}
\ddot q_j
= -U_j'(q_j)+\sum_{i\in\GG:\,i\volna j}V_{ij}'(q_{i}-q_j) 
+ \de_{j\La}(K_j q_j + \thet_j), \qu j\in\GG.
\eee
Due to \rprop\ref{th:ssupp}(1), we have $\ssupp\hat q_j\subset\{0\}$ for any $j\in\La$.
Then, \rlem\ref{lem:der} implies that all time-derivatives of $q_j$
belong to the space $\CC_0$, that is
\bee\lbl{q1dottozero}
q_j^{(l)}(t)\to 0 \ass t\to\pm\infty
\qu \mbox{for any}\qu 1\leq l \leq r+3
\qmb{and} \qu j\in\La,
\eee
were we recall that the functions $q_j$ are $C^{r+3}$-smooth and their derivatives belong to the space $\CC_b$ for any $j\in\GG$, due to Proposition~\ref{th:ssupp}(2).

The goal of this step is to show that convergence  \eqref{q1dottozero} holds for any $j\in\GG$.
If $\La=\GG$, we get this automatically, so we pass directly to Step 2. Below we assume that $\La\neq\GG$.
Due to \rprop\ref{th:ssupp}(2), 
we can differentiate equation \eqref{eq1} in time. 
We get
\bee\lbl{doteq}
\dddot q_j
= -U_j''(q_j)\dot q_j+\sum_{i\in\GG:\,i\volna j}V_{ij}''(q_{i}-q_j) (\dot q_{i}-\dot q_j)
+ \de_{j\La}(K_j \dot q_j + \dot\thet_j).
\eee
In view of convergence \eqref{q1dottozero},  
for $j\in\La$ the l.h.s. of equation \eqref{doteq} 
goes to zero as $t\to\pm\infty$, 
as well as all the terms from the r.h.s. which are multiplied by $\dot q_l$ with $l\in\La$,  and as well as the function $\dot\thet_j$. 
Then, the sum of remaining terms has to vanish as well, that is
\bee\lbl{sumVij}
\sum_{\substack{i\in\GG\sm\La:\\ i\volna j}}V_{ij}''(q_{i}-q_j) \dot q_{i} \to 0 \ass t\to\pm\infty \qmb{for any} \qu j\in\La.
\eee
By construction of the set $\La_\Ga$ and in view of assumption A5, 
there exists $k\in\La$ for which there is a unique $n\in\GG\sm\La$ satisfying $n\volna k$. 
Then, the convergence \eqref{sumVij} implies
\bee\lbl{V''q_1}
V_{nk}''(q_n-q_{k})\dot q_{n}(t)\to 0 \ass t\to\pm\infty.
\eee
In Step 3 of the proof we will show that convergence \eqref{V''q_1} implies 
\bee\lbl{q2tozero}
\dot q_n(t) \to 0 \ass t\to\pm\infty.
\eee
Now let us assume that \eqref{q2tozero} takes place,
so that $\dot q_n\in\CC_0$ and, consequently, 
$\ssupp \FF(\dot q_n)=\emptyset$.
Then, due to Lemma \ref{lem:der}, we have $q_n^{(l)}\in\CC_0$ for any $1\leq l\leq r+3$,
so that
$q_n^{(l)}(t)\ra 0$ as $t\ra\pm\infty$, 
and we have the convergence \eqref{q1dottozero} with the set $\La$ replaced by the set $\La\cup\{n\}$.
Next we repeat the argument above with  $\La$ replaced by $\La\cup\{n\}$, and iterate the procedure. Finally, we get the convergence \eqref{q1dottozero} with the set $\La$ replaced by $\La_\Ga$. Since, by assumption A5, we have $\La_\Ga=\GG$, the convergence \eqref{q1dottozero} holds for any $j\in\GG$.

{\it Step 2.} 
Next we let $t\to\pm\infty$ in equation \eqref{eq1}. 
Due to the convergence  \eqref{q1dottozero} which holds for any $j\in\GG$,
the functions $\ddot q_j$ vanish as $t\to\pm\infty$, as well as the functions $\thet_j$.
Then, $q(t)$ approaches a set consisting of points $q\in\mR^{|\GG|}$ 
satisfying
$$
0=-U_j'(q_j)+\sum_{i\in\GG:\,i\volna j}V_{ij}'(q_{i}-q_j) 
+ \de_{j\La}K_jq_j,
\qu j\in\GG.
$$
That is, we have $\dist(q(t),\NN^{\eff})\to 0$ as $t\to\pm\infty$, 
where $\NN^{\eff}$ is the set of critical points of the effective potential \eqref{effpot}.
Clearly, the latter convergence implies the desired convergence \eqref{main}.

{\it Step 3.} Finally, we deduce the convergence \eqref{q2tozero} from \eqref{V''q_1}.
Assume that \eqref{q2tozero} is false when $t\to +\infty$; the case $t\to-\infty$ can be studied similarly. 
Then there is $M>0$
such that there exists arbitrarily large time $t_0$
satisfying $|\dot q_n(t_0)|>M$. 
For definiteness we assume that 
$\dot q_n(t_0)$ is positive, so that
$\dot q_n(t_0)>M$.
Due to the uniform continuity of $\dot q_n$, 
there exists $\de>0$ independent of $t_0$, such that 
\bee\lbl{q2M}
\dot q_n(t)> M/2
\qu \mbox{for any}\qu t\in U_\de(t_0),
\eee
where $U_\de(t_0)$ denotes a $\de$-neighbourhood of $t_0$.
Since $\dot q_k(t)\to 0$ as $t\to \infty$, 
the inequality \eqref{q2M} implies that for sufficiently large $t_0$
we have
\bee\lbl{V''eps}
\fr{d}{dt}(q_n-q_k)(t)>M/4 \qu\mbox{for any}\qu t\in U_\de(t_0).
\eee
Take $\eps>0$ and choose $t_0$ to be so large that 
$
\big| V_{nk}''(q_n-q_k)\dot q_n\big|(t_0)<\eps M/2.
$
Then, by \eqref{q2M}, we have 
\bee\non
\big| V_{nk}''(q_n-q_k)\big|(t)<\eps \qu
\mbox{for all} \qu t\in U_\de(t_0).
\eee
Letting $\eps$ go to zero, we see that this 
contradicts to \eqref{V''eps}.
Indeed, by assumption A1, 
the function $V_{nk}''$ has only finite number of zeros on bounded sets,
while the function $(q_n-q_k)(t)$ is bounded.
\qed  

\subsection{Proof of Corollary \ref{lem:entr}}
\lbl{sec:entr}

Applying \rprop\ref{th:ssupp}(3)
to the formula \eqref{energy0},
we obtain
\begin{align*}
E(q_0,\xi_0,\dot q_0, \dot\xi_0)
&=E(q,\xi,\dot q, \dot\xi)(t) \\
&=\sli_{j\in\GG} \fr{\dot q_j(t)^2}{2}+ \sli_{m\in\La}\EE_m(q,\xi,\dot q, \dot\xi)(t) + V^{\eff}(q(t)) -  \sli_{m\in\La}\fr{K_mq_m(t)^2}{2}
+ \thet(t),
\end{align*}
where the function $\thet$ belongs to the space $\CC_0$.
Then, using  \eqref{dertozero} and \eqref{q_c}, 
we get the desired convergence \eqref{sumen}.

It is straightforward to check that,
due to equation \eqref{eq2},
$\dot\EE_m=q_m\ilif\ka_m\dot\xi_m\,d\nu$.
Then, using \rprop\ref{th:ssupp}(3), we find
$$
\dot\EE_m=K_mq_m\dot q_m + \dot\thet_m,
$$
where the function $\thet_m$ is $C^{r+1}$-smooth 
and belongs to the space $\CC_0$ together with all its derivatives.
Then   \eqref{deren} follows from \eqref{dertozero}.
\qed

\section{Energy transport in the system of one oscillator coupled to two thermostats}
\lbl{sec:oo}

In this section we consider a system of
one oscillator interacting with two thermostats, given by the equation
\bee\lbl{feq}
\ddot q = -U'(q) + \ilif \ka\xi_1\,d\nu + \ilif \ka\xi_2\,d\nu,
\quad
\ddot\xi_m(\nu)=-\nu^2\xi_m(\nu) + \ka(\nu) q,\qu m=1,2,
\eee
where 
$q(0)=q_0, \dot q(0)=\dot q_0$, $\xi_m(0)=\xi_{0m}$, $\dot\xi_m(0)=\dot\xi_{0m}$.
Our goal is to study the asymptotic behaviour of the energies $\EE_1, \,\EE_2$ of the thermostats,
which are defined as in \eqref{EEm}.
We assume that the functions $U$ and $\ka$ satisfy assumptions A1-A3 and A6
while the initial conditions fulfil assumption A4 with $m=1,2$.
Formally, the system  \eqref{feq} does not satisfy
assumptions of \rsec \ref{sec:setup} since the 
oscillator interacts with {\it two} thermostats. 
However, the theory developed in the paper can be applied to this system 
as well. 
Indeed, to see this it suffices to change the variables
\bee\lbl{trans}
\zeta:=\fr{\xi_1+\xi_2}{\sqrt{2}}, \quad
\eta:=\fr{\xi_1-\xi_2}{\sqrt{2}}.
\eee
Then equation \eqref{feq} takes the form
\bee\lbl{feq1}
\ddot q = -U'(q) + \sqrt{2}\ilif \ka\zeta\,d\nu,
\quad
\ddot\zeta(\nu)=-\nu^2\zeta(\nu) + \sqrt{2}\ka(\nu) q,\qu
\ddot\eta(\nu)=-\nu^2\eta(\nu).
\eee
Equation \eqref{feq1} describes a system which consists of an oscillator interacting with a thermostat
that satisfy assumptions of Section~\ref{sec:setup},
and of another independent thermostat. 
\bp\lbl{lem:f}
The energies $\EE_1(t),\EE_2(t)$ converge as $t\to\pm\infty$.
\ep
{\it Proof.} Set 
$$
\bm{\xi}_m:=\dot\xi_m + i\nu\xi_m,
\qu \bm{\zeta}:=\dot\zeta + i\nu\zeta, 
\qu \bm{\eta}:=\dot\eta+i\nu\eta,
$$
where $m=1,2$.
The equations of motion of the thermostats from \eqref{feq1} written in the variables $\eta,\,\zeta$ take the form
\bee\lbl{eqmot}
\dot {\bm{\zeta}} = i\nu \bm{\zeta} + \sqrt{2}\ka q,
\quad
\dot {\bm{\eta}} = i\nu \bm{\eta}.
\eee
The energy $\EE_1$ has the form
\bee\lbl{fE1}
\EE_1=\fr12\|\bm\xi_1\|_{L^2}^2 = \fr14\|\bm\zeta + \bm\eta\|_{L^2}^2 
=\fr14 \big(\|\bm\zeta\|^2_{L_2} + \|\bm\eta\|^2_{L_2} 
+ 2\Re\lan\bm\zeta, \bm\eta\ran_{L_2} \big),
\eee
where $\|\cdot\|_{L^2}$ and $\lan\cdot,\cdot\ran_{L^2}$ denote the standard
norm and scalar product in the space $L^2$.
Similarly,
\bee\lbl{fE2}
\EE_2 = \fr12\|\bm\xi_2\|_{L^2}^2
=\fr14\|\bm\zeta - \bm\eta\|_{L^2}^2
=\fr14 \big(\|\bm\zeta\|^2_{L_2} + \|\bm\eta\|^2_{L_2} 
- 2\Re\lan\bm\zeta, \bm\eta\ran_{L_2} \big).
\eee
Applying \rcor\ref{lem:entr} to the system given by the first two equations of
\eqref{feq1}  
we find that the energy of the thermostat $\zeta$,
given by  
$\|\bm\zeta\|^2_{L_2}$,
converges as $t\to\pm\infty$.
Moreover, the second equation from \eqref{eqmot} implies that the norm 
$\|\bm\eta\|_{L^2}$ is independent of time. 
Thus, in view of \eqref{fE1} and \eqref{fE2},
to finish the proof of the proposition it suffices to show that
the scalar product
$
\lan\bm\zeta, \bm\eta\ran_{L_2} 
$
converges as $t\to\pm\infty$.

Set $\bm\eta_0=\bm\eta(0)$ and $\bm\zeta_0=\bm\zeta(0)$.
Recall that $\bm{\zeta}_0, \bm{\eta}_0 \in L^2$,
due to Assumption A4.
In view of \eqref{eqmot}, we have
$$
\bm\eta=e^{i\nu t}\bm\eta_0, \quad
\bm\zeta = e^{i\nu t}\bm\zeta_0 
+\sqrt{2}\ka\ili_0^t e^{i\nu (t-\tau)}q(\tau)\,d\tau
=e^{i\nu t} (\bm\zeta_0 + \sqrt{2}\ka\hat q^t),
$$
where $q^t:=q \mI_{[0,t]}$ if $t\geq 0$, 
$q^t:=-q \mI_{[t,0]}$ if $t<0$ and $\hat q^t=\FF(q^t)$.
Set 
$q^\pm:=q\mI_{[0,\pm\infty)}$.
In \rlem\ref{lem:q+} it is 
shown  that $\ka \hat q^\pm\in L^2$ and
$\ka \hat q^t \to \ka \hat q^\pm$ weakly in $L^2$ as $t\to\pm\infty$. 
Then we have
\bee\lbl{fdeltaElim}
\lan\bm\eta, \bm\zeta\ran_{L_2}
=\lan\bm\eta_0, \bm\zeta_0 
+\sqrt{2}\ka\hat q^t\ran_{L_2}
\to 
\lan\bm\eta_0, \bm\zeta_0 
+\sqrt{2}\ka\hat q^{\pm}\ran_{L_2} \ass t\to\pm\infty.
\eee
\qed

\ssk
Now we study the question if we generically have  
\bee\lbl{nonsense}
\lim\limits_{t\to\star\infty}\EE_1(t)= \lim\limits_{t\to\star\infty}\EE_2(t), 
\eee
where $\star=+$ or $\star=-$.
Equations \eqref{fE1}-\eqref{fE2} imply that the equality \eqref{nonsense} takes place if and only if
we have $\lim\limits_{t\to\star\infty}\Re\lan\bm\eta,\bm\zeta\ran=0$.  
In turn, due to \eqref{fdeltaElim}, the latter convergence takes place if and only if 
\bee\lbl{nonsenser}
\Re\lan\bm\eta_0, \bm\zeta_0 
+\sqrt{2}\ka\hat q^{\star}\ran_{L_2}=0.
\eee
Obviously, the function $\ka\hat q^{\star}$ depends on the initial conditions $q_0, \dot q_0$, so that the equality $\bm\zeta_0 
+\sqrt{2}\ka\hat q^{\star}\equiv 0$
is generically false.
Moreover, the function $\bm\zeta_0 
+\sqrt{2}\ka\hat q^{\star}$
does not depend on the initial conditions $\bm\eta_0$.
Consequently, the equality \eqref{nonsenser} holds only for a set of the initial conditions 
$q_0,\dot q_0,\bm\zeta_0, \bm\eta_0$ of codimension one, so that \eqref{nonsense} is not generic.
For example, \eqref{nonsense} takes place if $\bm\eta_0=0$,
that is when we have $(\xi_{01},\dot\xi_{01})=(\xi_{02},\dot\xi_{02})$, so that the thermostats have the same initial conditions.

\bigskip
{\bf Acknowledgments.} I am grateful to No\'e Cuneo and Dmitry Treschev for useful discussions.
This work is supported by the Russian Science Foundation under grant 14-50-00005.

\appendix
\numberwithin{equation}{section}


\begin{thebibliography}{0}
\bibitem{BeKLeLu} C. Bernardin, V. Kannan, J.L. Lebowitz, J. Lukkarinen, {\it Harmonic Systems with Bulk Noises}, Journal of Statistical Physics, {\bf 146} (2012),  800-831.
\bibitem{Bog} V.I. Bogachev, {\it Measure theory}, v. 1, Springer-Verlag, Berlin (2007).
\bibitem{BLL} F. Bonetto, J. L. Lebowitz, J. Lukkarinen, {\it Fourier's Law for a Harmonic Crystal with Self-Consistent Stochastic Reservoirs}, Journal of Statistical Physics, {\bf 116} (2004),  783-813.
\bibitem{BLLO} F. Bonetto, J. L. Lebowitz, J. Lukkarinen, S. Olla, {\it Heat Conduction and Entropy Production in Anharmonic Crystals with Self-Consistent Stochastic Reservoirs}, Journal of Statistical Physics, {\bf 134} (2009), 1097-1119.  
\bibitem{Car} P. Carmona, {\it Existence and uniqueness of an invariant measure for a chain of oscillators in contact with two heat baths}, Stochastic Process. Appl., {\bf 117} (2007), 1076-1092.
\bibitem{CE} N. Cuneo, J.-P. Eckmann, {\it Non-equilibrium steady states for chains of four rotors},  Commun. Math. Phys., {\bf 345} (2016), 185-221.
\bibitem{CEHRB} N. Cuneo, J.-P. Eckmann, M. Hairer, L. Rey-Bellet, {\it Non-Equilibrium Steady States for Networks of
	Oscillators}, arXiv:1712.09413.  
\bibitem{CEPo} N. Cuneo, J.-P. Eckmann, C. Poquet, {\it Non-equilibrium steady state and subgeometric ergodicity for a chain of three coupled rotors}, Nonlinearity, {\bf 28} (2015), 2397-2421.
\bibitem{CPo} N. Cuneo, C. Poquet, {\it On the relaxation rate of short chains of rotors interacting with Langevin thermostats}, ECP {\bf 22} (2017), 1-8.
\bibitem{Dym12} A.V. Dymov, {\it Dissipative effects in a linear Lagrangian system with infinitely many degrees of freedom}, Izv. Math.  {\bf 76} (2012), 1116-1149.
\bibitem{Dym15} A. Dymov, {\it Nonequilibrium Statistical Mechanics of Hamiltonian Rotators with Alternated Spins}, 
Journal of Statistical Physics, {\bf 158} (2015), 968-1006.
\bibitem{Dym16} A. Dymov, {\it Nonequilibrium Statistical Mechanics of Weakly Stochastically Perturbed System of Oscillators},  Ann. IHP (A) {\bf 17} (2016), 1825-1882. 
\bibitem{Dym16a} A. Dymov, {\it Nonequilibrium statistical mechanics of a solid in medium}, Proc. Steklov Inst. Math., {\bf 295} (2016), 95-128.
\bibitem{EPRB} J.-P. Eckmann, C.-A. Pillet, L. Rey-Bellet, {\it Non-equilibrium statistical mechanics of anharmonic chains coupled to two heat baths at different temperatures}, Commun. Math. Phys., {\bf 201} (1999), 657-697.
\bibitem{EPRBa} J.-P. Eckmann, C.-A. Pillet and L. Rey-Bellet, {\it Entropy production in nonlinear, thermally driven Hamiltonian systems}, J. Stat. Phys., {\bf 9} (1999), 305-331.
\bibitem{GT} E.A. Gorin, D.V. Treschev, {\it Relative Version of the Titchmarsh Convolution Theorem}, Funct. Anal. Appl., {\bf 46} (2012), 26-32.
\bibitem{Kha} R. Khasminskii, {\it Stochastic stability of differential equations}; 2nd ed., Springer-Verlag, Berlin (2012).
\bibitem{Ko_r} A. Komech, {\it Attractors of nonlinear Hamilton PDEs}, Discrete Contin. Dyn. Syst., {\bf 36} (2016), 6201-6256.
\bibitem{KSK} A. Komech, M. Kunze, H. Spohn, {\it Long-Time asymptotics for a 
classical particle interacting with a scalar wave field},
Comm. Partial Differential Equations, {\bf 22} (1997), 307-335.
\bibitem{RBT} L. Rey-Bellet,  L.E. Thomas, {\it Exponential convergence to non-equilibrium stationary states in classical statistical mechanics}, Commun. Math. Phys., {\bf 225} (2002),  305-329.
\bibitem{Sau} S.M. Saulin, {\it Dissipation effects in infinite-dimensional Hamiltonian systems}, Theoret. and Math. Phys., {\bf 191} (2017), 537-557.
\bibitem{Tr} D. Treschev, {\it Oscillator and thermostat}, Discrete Contin. Dyn. Syst., {\bf 28}  (2010), 1693-1712.
\bibitem{Ver87} A. Veretennikov, {\it Bounds for the Mixing Rate in the Theory of Stochastic Equations}, Theory Probab. Appl., {\bf 32}  (1987), 273-281.
\bibitem{Vl} V.S. Vladimirov, {\it Generalized functions in mathematical physics}, Mir, Moscow (1979).
\end{thebibliography}
\end{document}